\documentclass[10pt,english]{article}
\usepackage[T1]{fontenc}
\usepackage[utf8]{inputenc}
\usepackage{geometry}
\usepackage{mathtools}
\usepackage{amsmath}
\usepackage{amssymb}
\usepackage{graphicx}
\usepackage{esint}

\makeatletter

\usepackage{bbm}

\usepackage[english]{babel}
\usepackage{amsfonts}\usepackage{mathrsfs}\usepackage{amsthm}\usepackage{bm}\usepackage{color}
\usepackage{dsfont}

\usepackage{amsfonts}
\usepackage{color}\usepackage{epsfig}
\usepackage{lscape}
\usepackage{footnote}
\usepackage{amsthm}
\usepackage{wasysym}\usepackage{amsfonts}
\usepackage[english]{babel}
\usepackage{tikz}
\usepackage{empheq}

\tikzset{square arrow/.style={to path={-- ++(0,-.25) -| (\tikztotarget)}}}

\allowdisplaybreaks[1]

\DeclareMathOperator{\sym}{sym}
\DeclareMathOperator{\skw}{skew}
\DeclareMathOperator{\tr}{tr}

\DeclareMathOperator{\anti}{anti}

\newcommand{\id}{{\boldsymbol{\mathbbm{1}}}}

\def\curl{\textrm{curl\,}}

\def\id{1\!\!1}

\let\@fnsymbol\@arabic

\makeatother

\usepackage{babel}
\begin{document}

\title{Correct traction boundary conditions  in the indeterminate couple stress model}

\author{\normalsize{Patrizio Neff\thanks{ Patrizio Neff,  \ \ Head of Lehrstuhl f\"{u}r Nichtlineare Analysis und Modellierung, Fakult\"{a}t f\"{u}r
Mathematik, Universit\"{a}t Duisburg-Essen,  Thea-Leymann Str. 9, 45127 Essen, Germany, email: patrizio.neff@uni-due.de}\quad
and \quad Ionel-Dumitrel Ghiba\thanks{Ionel-Dumitrel Ghiba, \ \ \ \ Lehrstuhl f\"{u}r Nichtlineare Analysis und Modellierung, Fakult\"{a}t f\"{u}r Mathematik,
Universit\"{a}t Duisburg-Essen, Thea-Leymann Str. 9, 45127 Essen, Germany;  Alexandru Ioan Cuza University of Ia\c si, Department of Mathematics,  Blvd.
Carol I, no. 11, 700506 Ia\c si,
Romania; and  Octav Mayer Institute of Mathematics of the
Romanian Academy, Ia\c si Branch,  700505 Ia\c si, email: dumitrel.ghiba@uni-due.de, dumitrel.ghiba@uaic.ro}  \quad
and \quad
Angela Madeo\footnote{Angela Madeo, \ \  Laboratoire de G\'{e}nie Civil et Ing\'{e}nierie Environnementale,
Universit\'{e} de Lyon-INSA, B\^{a}timent Coulomb, 69621 Villeurbanne
Cedex, France; and International Center M\&MOCS ``Mathematics and Mechanics
of Complex Systems", Palazzo Caetani,
Cisterna di Latina, Italy,
 email: angela.madeo@insa-lyon.fr}\quad and \quad Ingo M\"unch\thanks{Ingo M\"unch, Institute for Structural Analysis, Karlsruhe Institute of Technology, Kaiserstr. 12, 76131 Karlsruhe,
Germany, email: ingo.muench@kit.edu}}
}
\maketitle
\begin{abstract}
In this paper we consider the Grioli-Koiter-Mindlin-Toupin indeterminate
couple stress model. The main aim is to show that the traction boundary
conditions were not yet completely deduced. As it turns out, and
to our own surprise, restricting the boundary condition framework
from the strain gradient models to the couple stress model does not
reduce to  Mindlin's set of accepted boundary
conditions. We present therefore, for the first time the complete,
consistent set of traction boundary conditions. \\
 \vspace*{0.25cm}
\end{abstract}

\section{Introduction}

Higher gradient elasticity models are nowadays increasingly used to
describe mechanical structures at the micro- and nano-scale or to
regularize certain ill-posed problems by means of higher gradient
contributions \cite{Mindlin65,Mindlin68}. One of the very first among
such models is the so called indeterminate couple stress model \cite{Grioli60,Mindlin62,Toupin64,Koiter64}
in which the higher gradient contributions only enter through gradients
on the continuum rotation, i.e. the total elastic energy can be written
as $W(\nabla u,\nabla(\nabla u))=W_{e}({\rm sym}\nabla u)+W_{{\rm curv}}(\nabla({\rm curl}u))$.

The question of boundary conditions in higher gradient elasticity
models has been a subject of continuous attention. The crux of the
matter in higher gradient models is the impossibility to vary the
test function and its gradient independently. A suitable split into
tangential and normal parts must always be considered. This is well
known in general higher gradient models, see e.g. \cite{bleustein1967note,TierstenBleustein}.
The indeterminate couple stress model has been investigated in this
respect as well. A first answer has been given by Mindlin and Tiersten
\cite{Mindlin62} as well as Koiter \cite{Koiter64} who established
(correctly) that only 5 geometric and 5 traction boundary conditions
can be prescribed due to the dependence of the curvature energy only
on gradients of rotations. We agree that there are 5 traction boundary
conditions in the indeterminate couple stress model which may be independently
prescribed. However, we show in \cite{MadeoGhibaNeffMunchKM,NeffGhibaMadeoMunch}
that the correct traction boundary conditions are not those proposed
by Mindlin and Tiersten \cite{Mindlin62} and which are currently
used in the literature. Since all papers dealing with the indeterminate
couple stress model use this incomplete set of boundary conditions
we will not refer further to any specific one.

\section{The indeterminate couple stress model}

We consider a body which occupies a bounded open set $\Omega$ of
the three-dimensional Euclidian space $\mathbb{R}^{3}$ and assume
that its boundary $\partial\Omega$ is a piecewise smooth surface.
An elastic material fills the domain $\Omega\subset\mathbb{R}^{3}$
and we refer the motion of the body to rectangular axes $Ox_{i}$,
$i=1,2,3$. For vector fields $v$ with components $v_{i}\in{\rm H}^{1}(\Omega)$,
$i=1,2,3$, we define $\nabla\, v=\left((\nabla\, v_{1})^{T},(\nabla\, v_{2})^{T},(\nabla\, v_{3})^{T}\right)^{T}$,
while for tensor fields $P$ with the rows $P_{i}\in{\rm H}({\rm div}\,;\Omega)$,
$i=1,2,3$, we define ${\rm Div}\, P=\left({\rm div}\, P_{1},{\rm div}\, P_{2},{\rm div}\, P_{3}\right)^{T}.$ Equivalently, in index notation: $(\nabla v)_{ik}=v_{i,k}$ and $({\rm Div}\, P)_i=P_{ij,j}$.
In the remainder of the paper, $\sym X$ and $\skw X$ denote the symmetric and the skew symmetric
part of the matrix $X$, respectively, $\tr(X)$ denotes the trace
of the matrix $X$, $\|X\|$ is the Frobenius norm of the matrix $X$. The identity tensor on $\mathbb{R}^{3\times3}$ will
be denoted by $\id$. We also use the operator $\anti:\mathbb{R}^{3}\rightarrow\mathfrak{so}(3)$,
$\mathfrak{so}(3):=\{X\in\mathbb{R}^{3\times3}\;|X^{T}=-X\}$, defined
by $(\anti(v))_{ij}=-\varepsilon_{ijk}v_{k},\forall\, v\in\mathbb{R}^{3}$,
where $\varepsilon_{ijk}$ is the totally antisymmetric third order
permutation Levi-Civita tensor. We use the $\curl$ operator, $\curl v=\varepsilon_{ijk}v_{k,j},\forall\, v\in\mathbb{R}^{3}$
and denote respectively by $\cdot\:$, $:$ and $\left\langle \cdot,\cdot\right\rangle $
a simple and double contraction and the scalar product between two
tensors of any suitable order%
\footnote{For example, $(A\cdot v)_{i}=A_{ij}v_{j}$, $(A\cdot B)_{ik}=A_{ij}B_{jk}$,
$A:B=A_{ij}B_{ji}$, $(C\cdot B)_{ijk}=C_{ijp}B_{pk}$, $(C:B)_{i}=C_{ijp}B_{pj}$,
$\left\langle v,w\right\rangle =v\cdot w=v_{i}w_{i}$, $\left\langle A,B\right\rangle =A_{ij}B_{ij}$
etc. %
}. Everywhere we adopt the Einstein convention of sum over repeated
indices if not differently specified.

The Grioli-Koiter-Mindlin-Toupin isotropic indeterminate couple stress
model \cite{Grioli60,Mindlin62,Toupin64,Koiter64} considers the curvature
energy $W_{{\rm curv}}(\nabla(\curl u))=\frac{\alpha_{1}}{4}\,\|\sym\nabla\curl\, u\|^{2}+\frac{\alpha_{2}}{4}\,\|\skw\nabla\curl\, u\|^{2}$
and the classical elastic energy $W_{e}(\sym\nabla u)=\mu\,\|{\rm sym}\nabla u\|^{2}+\frac{\lambda}{2}\,[\tr({\rm sym}\nabla u)]^{2},$
where $\mu,\lambda,\alpha_{1}$ and $\alpha_{2}$ are constitutive
coefficients. The correct and accepted strong form of the Euler-Lagrange
equations are
\begin{equation}
\begin{array}{rll}
{\rm Div}\!\! & (\sigma-\frac{1}{2}\,\anti({\rm Div}\,\widetilde{m}))+f=0,&\text{equilibrium of forces}\\
\sigma & =D_{\sym\nabla u}W_{e}(\sym\nabla u)=2\,\mu\,\sym\nabla u+\lambda\,\tr(\nabla u)\id,&\text{symmetric Cauchy-stress}\\
\widetilde{m} & =D_{\nabla\curl u}W_{{\rm curv}}(\nabla\curl u)=\alpha_{1}\,\sym(\nabla\curl u)+\alpha_{2}\,\skw(\nabla\curl u),&\text{couple stress tensor}.
\end{array}
\end{equation}
Note that the couple stress tensor $\widetilde{m}$ is a second order
and trace free tensor. Having the Euler-Lagrange equation, the question
of which boundary conditions may be prescribed arises.

\section{The incomplete boundary conditions considered in literature}

We want to stress the fact that in the framework of a complete second
gradient theory we can arbitrarily prescribe $u$ and the normal derivative
of the displacement $\nabla u\cdot n$ on the Dirichlet boundary $\Gamma$.
This means that one has $6$ independent geometric (or kinematical)
boundary conditions that can be assigned on the boundary of the considered
second gradient medium. Analogously, one can assign $6$ traction
(or natural) conditions on the force (in duality of $u$) and double
force (in duality of $\nabla u\cdot n$), respectively, at $\partial\Omega\setminus\overline{\Gamma}$.
The situation is slightly different in the indeterminate couple stress
model since only a certain linear combination of second derivatives,
i.e. $\nabla\curl u$, is controlled. Mindlin and Tiersten \cite{Mindlin62}
concluded that the geometric boundary conditions on $\Gamma\subset\partial\Omega$
are the five independent conditions
\begin{equation}
\begin{array}{rl}
u\Big|_{\Gamma} & =\ \widetilde{u}^{0},\qquad(\id-n\otimes n)\cdot\curl u\Big|_{\Gamma}=(\id-n\otimes n)\cdot\curl\widetilde{u}^{0},\end{array}\label{bcme1}
\end{equation}
for a given vector function $\widetilde{u}^{0}$ at the boundary,
where $n$ is the unit normal vector on $\partial\Omega$ and $\otimes$
denotes the dyadic product of two vectors. The latter condition, in
fact, prescribes only the tangential component of $\curl u$. Therefore,
one may prescribe only 5 independent boundary conditions.

The possible traction boundary conditions on the remaining boundary
$\partial\Omega\setminus\overline{\Gamma}$ given first by Mindlin
and Tiersten \cite{Mindlin62} are
\begin{equation}
\begin{array}{rl}
\left\{ \left(\sigma-\frac{1}{2}\,\anti({\rm Div}\,\widetilde{m})\right)\cdot\, n-\frac{1}{2}n\times\nabla[\langle n,(\sym\widetilde{{m}})\cdot n\rangle]\right\} \Big|_{\partial\Omega\setminus\overline{\Gamma}} & =\widetilde{t},\\
(\id-n\otimes n)\cdot \widetilde{{m}}\cdot n\Big|_{\partial\Omega\setminus\overline{\Gamma}} & =(\id-n\otimes n)\cdot\widetilde{g},
\end{array}\label{bcme2}
\end{equation}
for prescribed vector functions $\widetilde{t}$ and $\widetilde{g}$
at the boundary, where $\langle\cdot,\cdot\rangle$ denotes the scalar
product of two vectors. Mindlin and Tiersten \cite{Mindlin62} have
correctly concluded that the maximal number of independent traction
boundary conditions is also 5. The same conclusion has been arrived
at by Koiter \cite{Koiter64}. These traction boundary conditions
(\ref{bcme2}) have been rederived again and again. However, they
are erroneous.

\section{The correct boundary conditions in the indeterminate couple stress
model}

The prescribed traction boundary conditions (\ref{bcme2}) proposed
by Mindlin and Tiersten \cite{Mindlin62} do not remain independent,
in the sense that $\widetilde{g}$ leads to a further energetic conjugate,
besides $\widetilde{t}$, of $u$. From this reason and looking back
to the clear and correct boundary conditions considered in the more
general second gradient elasticity model, in order to prescribe independent
geometric boundary conditions and their corresponding completely independent
energetic conjugate (traction boundary conditions), we have to prescribe
$u$ and $(\id-n\otimes n)\cdot\left(\nabla u\cdot n\right)$. Let
us remark that prescribing $u\big|_{\Gamma}=\widetilde{u}^{0}$ and
$(\id-n\otimes n)\cdot\left(\nabla u\cdot n\right)\big|_{\Gamma}=(\id-n\otimes n)\cdot\nabla\widetilde{u}^{0}\cdot n$
is fully equivalent with prescribing $u\big|_{\Gamma}=\widetilde{u}^{0}$
and $(\id-n\otimes n)\cdot\curl u\big|_{\Gamma}=(\id-n\otimes n)\cdot\curl\widetilde{u}^{0}$,
which is (\ref{bcme1}). However, in the formulation of the principle
of virtual power, the energetic conjugate of $(\id-n\otimes n)\cdot\curl u$
is not equal to the energetic conjugate of $(\id-n\otimes n)\cdot\left(\nabla u\cdot n\right)$.

Using the principle of virtual power proposed by Mindlin and Tiersten
\cite[Eq. (5.13)]{Mindlin62},  but now suitably applying the surface
divergence theorem \cite{MadeoGhibaNeffMunchKM,dell2012beyond,dell2012contact}, we arrive at the
following traction boundary conditions on $\partial\Omega\setminus\overline{\Gamma}$
\begin{equation}
\hspace{-0.1cm}\begin{array}{lll}
\Big\{\Big(\sigma & -\frac{1}{2}\,\anti({\rm Div}\,\widetilde{m})\Big)\cdot n-\frac{1}{2}n\times\nabla[\langle n,(\sym\widetilde{{m}})\cdot n\rangle] & \ \vspace{1.2mm}\\
 & -\frac{1}{2}\nabla\left[\:\anti\left(\:(\id-n\otimes n)\cdot\widetilde{{m}}\cdot n\:\right)\cdot(\id-n\otimes n)\:\right]:(\id-n\otimes n)\Big\}\Big|_{\partial\Omega\setminus\overline{\Gamma}} & =\widetilde{t},\\
 & \hspace{3.7cm}(\id-n\otimes n)\cdot\anti[(\id-n\otimes n)\cdot\widetilde{{m}}\cdot n]\cdot n\Big|_{\partial\Omega\setminus\overline{\Gamma}} & =(\id-n\otimes n)\cdot\widetilde{g},
\end{array}\label{alter1}
\end{equation}
together with the traction boundary conditions on $\partial{\Gamma}$
\begin{equation}
\begin{array}{rl}
\left\{ ([\anti[(\id-n\otimes n)\cdot\widetilde{{m}}\cdot n]]^{+}-[\anti[(\id-n\otimes n)\cdot\widetilde{{m}}\cdot n]]^{-})\cdot\nu\right\} \Big|_{\partial{\Gamma}} & =\widetilde{\pi},
\hspace{2.5cm}
\end{array}\label{alter2}
\end{equation}
where $\widetilde{t}$ and
$\widetilde{g}$ are prescribed vector functions on $\partial\Omega\setminus{\overline{\Gamma}}$,
while $\widetilde{\pi}$ is a prescribed vector function on $\partial\Gamma$.
Here, $\nu$ is a vector tangential to the surface $\Gamma$ and which
is orthogonal to its boundary $\partial\Gamma$. The term $[\anti[(\id-n\otimes n)\cdot\widetilde{{m}}\cdot n]]^{+}-[\anti[(\id-n\otimes n)\text{\ensuremath{\cdot}}\widetilde{{m}}\cdot n]]^{-}$
measures the discontinuity of $\anti[(\id-n\otimes n)\cdot\widetilde{{m}}\cdot n]$
across $\partial\Gamma$.

Comparing (\ref{bcme2}) and (\ref{alter1}), we remark that in the
Mindlin and Tiersten formulation (\ref{bcme2})$_{2}$ it remains
a missing boundary term $-\frac{1}{2}\nabla\left[\:\anti\left(\:(\id-n\otimes n)\cdot\widetilde{{m}}\cdot n\:\right)\cdot(\id-n\otimes n)\:\right]:(\id-n\otimes n)$
which also performs work against $u$. On the other hand, we show in \cite{MadeoGhibaNeffMunchKM} that when
the higher gradient contributions only enter through gradients on
the continuum rotation, i.e., $\nabla \curl u$, the independent traction
boundary conditions which are coming from the representation in terms
of a particular case of second gradient elasticity model written with
third order moment tensors coincide with our novel traction boundary
conditions (\ref{alter1}) and (\ref{alter2}), and not with the traction
boundary conditions (\ref{bcme2}) proposed by Mindlin and Tiersten.

Our renewed interest in traction boundary conditions in the indeterminate
couple stress model was triggered by the controversial papers \cite{hadjesfandiari2011couple,hadjesfandiari2013fundamental}.
There, the authors have made far reaching claims on the possible anti-symmetric
nature of the second order couple stress tensor $\widetilde{m}$.
Their reasoning is based on physically plausible assumptions similar
to a Cosserat or micromorphic theory \cite{NeffGhibaMicroModel} which
led them to require a total split of the effect of force and moment
tensors in (\ref{bcme2}). In (\ref{bcme2}), this can be achieved
if and only if $\widetilde{m}$ is skew-symmetric and this constitutes
the essence of their claim. However, since (\ref{bcme2}) is incomplete,
their conclusion is misleading, see also \cite{NeffGhibagegenHD}.
The couple stress tensor in the indeterminate couple stress theory
is not necessarily skew-symmetric! Quite to the contrary, the couple
stress tensor may be chosen to be symmetric \cite{MunchYang,Neff_Jeong_IJSS09}.

\begin{footnotesize}

\end{footnotesize}
\end{document}